\documentclass[copyright]{eptcs}
\providecommand{\event}{MOD* 2014}
\usepackage{breakurl}

\usepackage{amssymb}
\usepackage{amsmath}
\usepackage{listings}
\usepackage{courier}
\usepackage{hyperref}
\usepackage[dvips]{graphicx}

\providecommand{\doi}[1]{\textsc{doi}: \href{http://dx.doi.org/#1}{\nolinkurl{#1}}}

\title{A Language Support for Exhaustive Fault-Injection in Message-Passing System Models}
\author{Masaya Suzuki \qquad\qquad Takuo Watanabe%
\institute{Department of Computer Science,\, Tokyo Institute of Technology,\, Tokyo, Japan}
\email{draftcode@psg.cs.titech.ac.jp \qquad\qquad takuo@acm.org}}

\def\titlerunning{A Language Support for Exhaustive Fault-Injection in Message-Passing System Models}
\def\authorrunning{M. Suzuki and T. Watanabe}

\begin{document}
\maketitle

\begin{abstract}
This paper presents an approach towards specifying and verifying adaptive distributed systems.
We here take fault-handling as an example of adaptive behavior and propose a modeling language Sandal for describing fault-prone message-passing systems.
One of the unique mechanisms of the language is a linguistic support for abstracting typical faults such as unexpected termination of processes and random loss of messages.
The Sandal compiler translates a model into a set of NuSMV modules.
During the compilation process, faults specified in the model will be woven into the output.
One can thus enjoy full-automatic exhaustive fault-injection without writing faulty behaviors explicitly.
We demonstrate the advantage of the language by verifying a model of the two-phase commit protocol under faulty environment.
\end{abstract}

\section{Introduction}\label{section:Introduction}

As large-scale computing is more prevalent than ever, adaptability in software systems gains increasing importance. In particular, context-awareness and dynamic self adaptability in changing environment is crucial for developing sustainable systems. In this paper, we take fault-handling as an example of dynamic adaptive behavior in faulty circumstances and propose a method for describing formal models of fault-prone message-passing systems. The proposal is a linguistic approach; we designed and developed a modeling language Sandal that provides abstractions of typical faults such as unexpected termination of processes and random loss of messages.

The Sandal compiler translates a model into a set of NuSMV\cite{Cimatti:2002aa} modules. During the translation process, faults specified in the model will be woven into the output; in other words, the compiler performs a sort of software fault-injection (SFI). SFI and model checking are both fundamental techniques for developing reliable software.  The combination of the both is a promising formal approach for verifying  software systems  in unreliable environments.

To model check a system, we usually describe the model of the system with a modeling language.
According to the levels of abstraction, we categorize modeling languages into low-level and high-level languages.
The former provides primitives for describing a model as a system of automata while the latter provides abstraction mechanisms for representing processes, channels, etc.
NuSMV\cite{Cimatti:2002aa} and LOTOS\cite{Bolognesi:1987aa} are examples of low-level languages, and Promela\cite{Holzmann:1997aa} and Rebeca\cite{Sirjani:2004aa} are categorized as high-level languages.

Both types of languages have their advantages and are used for modeling various systems. 
However, they too have difficulties in describing fault-prone systems due to the lack of proper mechanisms for expressing faulty behaviors.
For example, injecting a fault to a message reception part in a model may affect not only the part itself and its nearby parts but also other modules of the model.
These unwanted effects result in low modularity and maintainability of the model.

Automating fault-injection solves the above problem to some extent.
For this purpose, some automation tools such as MODIFI\cite{Svenningsson:2010aa} and FSAP/NuSMV-SA\cite{Bozzano:2003aa} have been developed.
As these tools are designed to treat hardware faults, they are not suitable for our purpose.
In addition, the above-mentioned modularity problem cannot be solved by these automation tools.

We adopt a linguistic approach to this problem and propose a high-level modeling language Sandal targeted to message-passing systems.
Sandal provides language constructs for specifying typical faults including timeout, unexpected termination of processes, random loss of messages.
The main advantage of this approach is the increased modularity of model descriptions owing to the language constructs.
Another advantage is that the injection of faults does not affect the original fault-free behaviors of the model in an unexpected way.

We give the semantics of Sandal as a translation to automata.
According to the semantics, we implemented a Sandal compiler that generates a set of NuSMV modules.
Every fault specified in the model is injected automatically by the compiler in a non-deterministic way so that all possible fault scenarios are generated on the fly by the method referred as \emph{exhaustive fault-injection}\cite{Steiner:2004aa}.

To demonstrate the advantage of our approach, we present a case study of two-phase commit protocol.
In the case study, we compare the models written in Sandal and Promela from the viewpoint of modularity and verification performance.
From a modularity viewpoint, Sandal achieves better performance with both the simplicity of specifying faults and the maintainability of models than Promela.
The verification result shows that Sandal can properly inject faults into the model and can verify the properties regarding the faults as expected.
The overhead of the verification speed of Sandal is in an acceptable range.

The rest of this paper is organized as follows.
The next section gives a detailed discussion of the problem.
Section~\ref{section:Sandal} presents the overview of our modeling language Sandal, and Section~\ref{section:CaseStudy} demonstrates its application to the two-phase commit protocol as a case study.
Section~\ref{section:FutureWork} mentions some future work and Section~\ref{section:Conclusion} concludes with a summary.

\section{Fault-Injection for Software Models}\label{section:Problems}

Injecting faults into a complex software model is sometimes a complicated process and requires manual instrumentation of the model.
Part of the reason is that software models are not the target of current automatic fault-injection tools.
They usually support the faults in a single variable like bit-flip or value-stuck, which are suitable for hardware models.
In contrast, typical fault in software systems involves multiple actions and/or variables; hence it cannot be expressed in an error of a single variable.

The cost of the manual instrumentation depends on the language used to describe the model.
If the model is constructed in a low-level modeling language that only provides the basic constructs representing state transitions in automata, users should describe the \emph{foundation layer} of a system as well as the system behaviors. 
Here, the foundation layer means the part of the model that the \emph{high-level} system model is built onto it.
For example, to describe a system that uses message-passing communication, we need to model the messaging mechanism and the scheduler using the basic language primitives.
On the other hand, if the model is constructed in a high-level modeling language that has the constructs corresponding to the foundation layer, we can easily build a system model on top of them.

Unfortunately, the both approaches incur some problems in modeling faults. 
In the low-level language approach, the distinction between the foundation layer and the system model is vague. This decreases the modularity of the model. To make things worse, the foundation and the injected faults sometimes affect the system model in an unexpected way. It is also the case that the primitives provided by the language is so inexpressive that faults cannot be implemented within the foundation layer. In this case, the fault descriptions ooze into the system model. The same problem may happen even if we use high-level languages. This kind of languages provide constructs regarding a foundation layer, but we cannot change its behavior.  This means that it is hard to implement the fault happened in the functionality provided by the foundation layer. It might be possible to overcome this restriction by adding an abstraction between a system model and the foundation, but it leads the same problem as the approach of the low-level languages.

\begin{figure}
  \centering
  \includegraphics[width=0.8\linewidth]{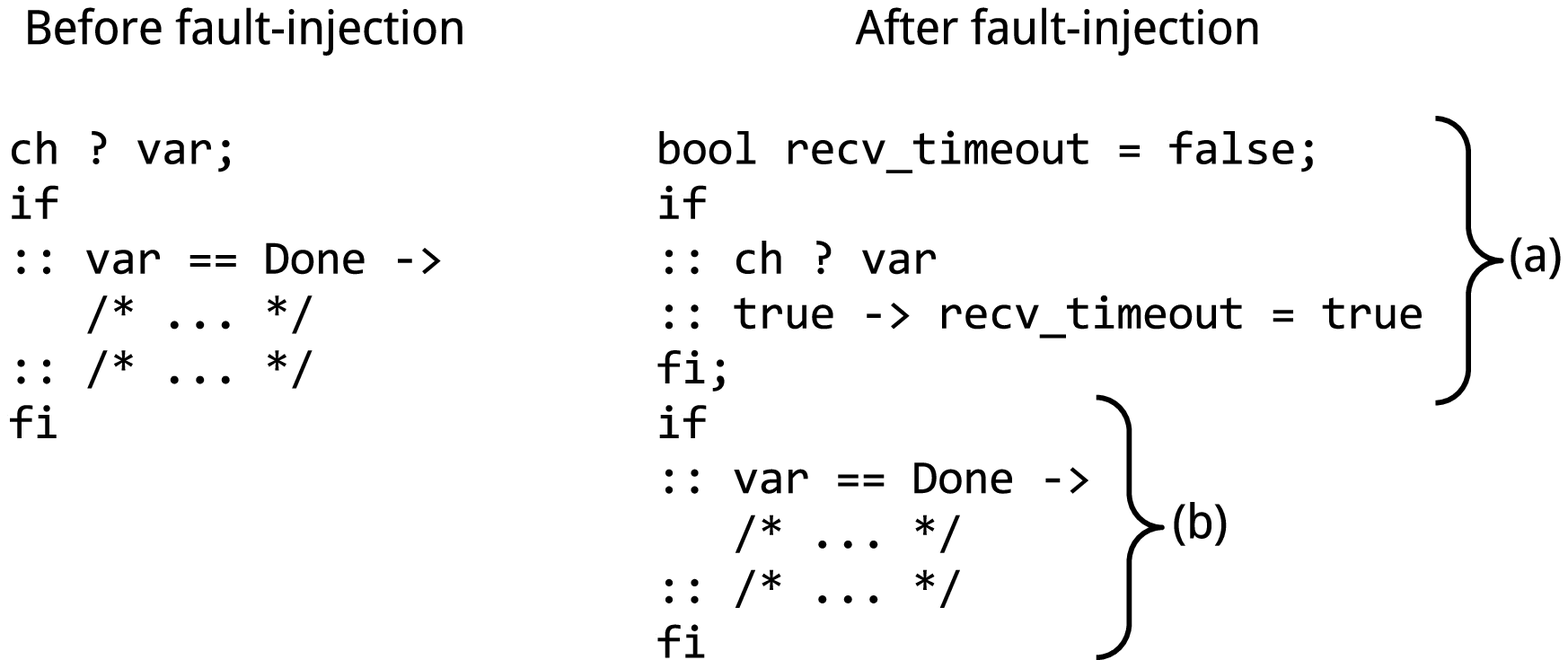}
  \caption{Implementing Fault-Injection in a Promela Model}
  \label{fault-injection-before-after.pdf}
\end{figure}

To illustrate the problems regarding fault-injection, we show two simple example models written in Promela. The first example is a timeout action in receiving messages. Due to the overload of network traffic, operations over the network sometimes take longer time than expected, and the system should give up the operation. Figure~\ref{fault-injection-before-after.pdf} shows an example of a timeout action implemented as fault-injection.  The original fault-free process receives a message from the channel \lstinline|ch| and does some computation based on the received value. 
In the fault-injected version of the model, the receive operation has a chance to timeout. The fault is implemented in the part (a). In this part, the timeout action is implemented as skipping the message reception.

Unlike other kinds of faults dealt with in this paper (random loss of messages and unexpected termination of processes), timeout faults are often explicitly handled in the models. This means that the part (b) of the model may treats the case in which the process fails to receive a message within an expected time.

\begin{figure}
  \centering
  \includegraphics{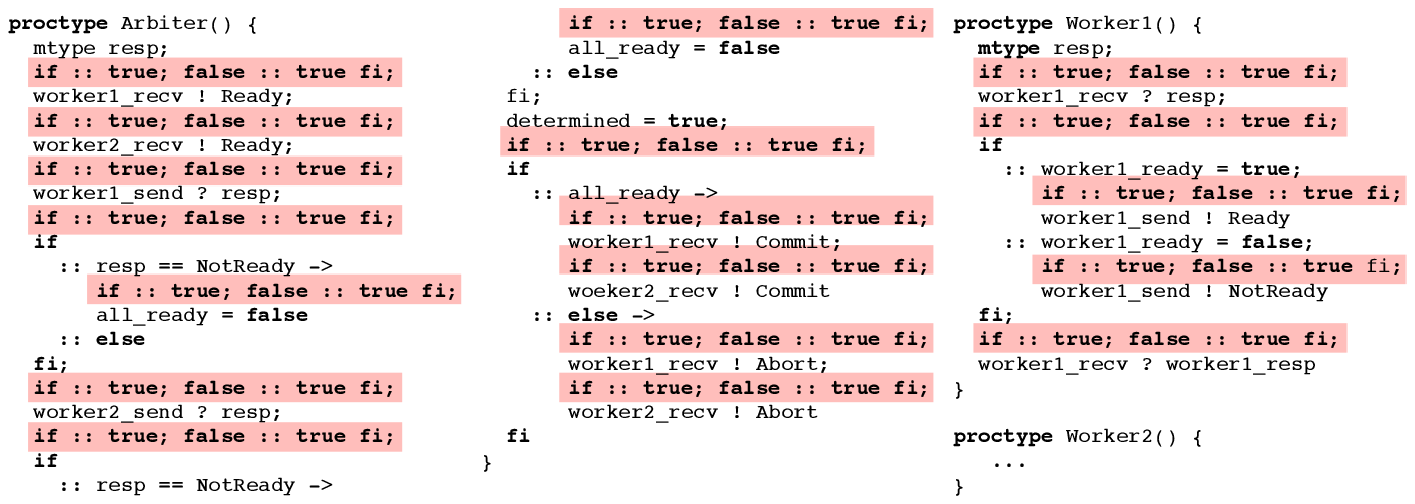}
  \caption{A Promela model with unexpected termination of processes}
  \label{manual-fault-implementation.pdf}
\end{figure}

Another example is the injection of unexpected termination faults.
This fault emulates unexpected shutdown or failure of machines. 
It can be described in the way that processes have a chance to stop its execution.
For example a simple Promela sentence
\begin{lstlisting}
        if :: true; false :: true fi
\end{lstlisting}
can model such behavior.
The problem is that the above sentence should be inserted into the place where each observable action happen.
Figure~\ref{manual-fault-implementation.pdf} shows the model after the fault-injection. The highlighted parts are the implementation of this fault.  They are scattered in the model, and this may doubles the size of the resulting model and decrease the maintainability.

\section{The Modeling Language Sandal}\label{section:Sandal}

\subsection{The Overview of the Language}

Sandal is designed to describe message passing systems. Message passing systems have processes communicating with each other only by message passing. In Sandal, a system consists of processes and channels. A process consists of a thread of execution and variables. Unlike the counterpart of real operating systems, it does not contain many threads. It has only one thread of execution. Sandal employs the shared-nothing model so processes cannot communicate via shared variables.

There are two types of channels in Sandal: rendezvous channels and buffered channels.  With rendezvous channels, two processes can communicate in a synchronous way; both a sender and a receiver should be ready on the same channel to communicate.  With buffered channels, two processes can communicate in an asynchronous way.  The values sent are saved in the buffer of a channel. If a receiver wants to receive a value, it tries to pop out from the buffer.

To achieve fully automatic exhaustive fault-injection, Sandal is aware of the faults. Although there are many faults that can be considered, Sandal treats three types of faults: unexpected termination of processes, random loss of messages, and timeout in receiving a message.

Unexpected termination of processes is a fault that processes are unintentionally shut down in arbitrary timing. This is intended to emulate the real situation that one of the machines in a distributed system is crashed.  Hardwares will, sooner or later, be broken. Even if one single computer has a low failure rate, the accumulated failure rate of the machines will be unignorable.  If such faults are recovered and abstracted away by hardware, it is all right with software. Unfortunately, there is the case that cannot be recovered by hardware; therefore, a system's state should be recovered by software. Thus, the fault tolerance property for unexpected termination of processes is one of the basic requirement for distributed systems.

Random loss of messages is a fault that some messages are randomly dropped when sending them. This occurs when a message is sent by an unreliable way like UDP.  Timeout in receiving a message is a fault that a transmission is not completed in some time window. They occur for various reasons: network overload, sending messages to the machine that is turned off, misconfiguration or maintenance of a router, and so on.
Such network unresponsiveness can be a temporary matter, so these faults may repeatedly appear. 
A possible solution is to send multiple copies of a message until the sender process receives the ACK for it.
However this protocol can not guarantee the reception of the message.

We implemented an experimental compiler that generates a set of NuSMV modules from a Sandal model.
The reason for employing NuSMV as the compilation target is that the semantics of a Sandal model can be expressed as a set of NuSVM modules in a straightforward manner.
The source code of the compiler (including some sample models) is available at the first author's GitHub repository\footnote{https://github.com/draftcode/sandal}.

\subsection{Syntax}

The syntax of Sandal is similar to that of the programming language Go\cite{golang-web}, 
which loosely follows the tradition of the programming language C.
Figure~\ref{sample.sandal} shows a simple model in Sandal that describes a system in which two processes exchange messages.

\begin{figure}
\begin{lstlisting}[basicstyle={\small\ttfamily},frame=lines]
proc Starter(recv_ch channel { bool }, send_ch channel { bool }) {
  var v bool
  send(send_ch, true); recv(recv_ch, v)
}
proc Receiver(recv_ch channel { bool }, send_ch channel { bool }) {
  var v bool
  recv(recv_ch, v); send(send_ch, true)
}
init {
  P0: Starter(receiver_to_starter, starter_to_receiver),
  P1: Receiver(starter_to_receiver, receiver_to_starter),
  receiver_to_starter: channel { bool },
  starter_to_receiver: channel { bool },
}
\end{lstlisting}
\caption{An Example Model in Sandal}\label{sample.sandal}
\end{figure}

\subsubsection{Process Templates}

A process definition, a construct starting with the keyword \lstinline|proc|, defines a template of processes.
The identifier after \lstinline|proc| is the name of the definition.  
In Figure~\ref{sample.sandal}, two templates named \lstinline|Starter| and \lstinline|Receiver| are defined.
A list of parameters follows after the template name.
The both template in the example have two parameters \lstinline|recv_ch| and \lstinline|send_ch| of 
type \lstinline|channel { bool }|, a channel type whose message contents are boolean values.
The last part of the template definition is a block (one or more statements wrapped in braces).

\subsubsection{Init-Blocks}

A block preceded by the keyword \lstinline|init| is called an \emph{init-block}.
It describes the configuration of processes and channels in the system to be defined.

\begin{figure}[tbp]
  \centering
  \includegraphics[width=0.8\linewidth]{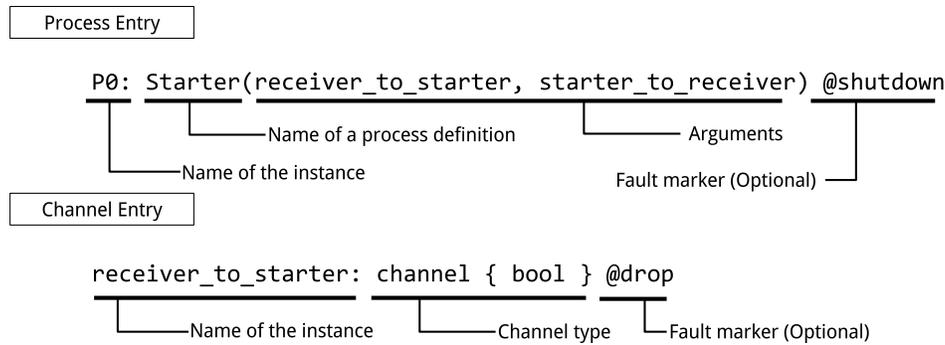}
  \caption{Two Types of Init-Block Entries}
  \label{init-block-entry.pdf}
\end{figure}

An init-block contains entries (called init-block entries) separated by commas. 
Each entry must be an instantiation of either a process or a channel. 
It starts with the name of the instance followed by a colon
and the rest part depends on what it represents (Figure~\ref{init-block-entry.pdf}).
If it is an instantiation of a process, the name of a process definition and its arguments follow.
If it is an instantiation of a channel, a channel type follows. 

\emph{Fault markers} can be attached to init-block entries.
A fault marker attached to a process (or a channel) states that the specified fault may occur in the process (or channel).
They are added to the last part of the entries in init-blocks.
The current version of Sandal provides two fault markers \lstinline|@shutdown| and \lstinline|@drop|.

\subsubsection{Messaging Statements}

Because the statement and expression syntax of Sandal closely matches that of traditional programming languages,
we only mention messaging statements in this subsection for brevity.
Statements \lstinline{recv} and \lstinline{peek} are used for receiving values from a channel. 
The difference is that \lstinline{recv} statements pop the values out from a channel while \lstinline{peek} statements just copy them. These operations have non-blocking and timeout variants.

The arguments to these messaging statements are treated specially. The first argument should be a channel. This is a channel that is used to communicate. The rest of the arguments should be variable names. After the statement is executed, the received values are stored to these variables.

\subsection{Semantics}

\subsubsection{Processes}

A process in a Sandal model can be seen as a state machine. 
Each transition in the machine may have a condition (called \emph{guard}) and side-effects (called {actions}).
For example, a guard may be \emph{``a value is ready to be received in the channel named c''} and an action may be \emph{``receive a value in the channel and store it in the variable named v.''}

\begin{figure}[h]
  \centering
  \includegraphics[scale=.6]{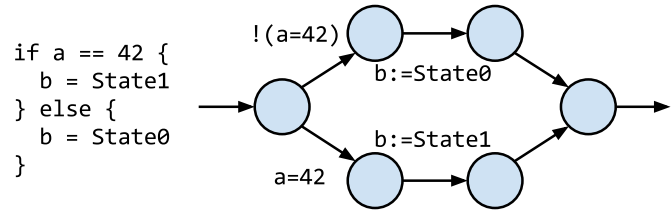}
  \caption{A Simple Statement and its Semantics}
  \label{process-semantics.pdf}
\end{figure}

A process is a graph of statements that are executed in order. 
For example, Figure~\ref{process-semantics.pdf} shows a composite (\lstinline{if}) statement and its semantics.
There are two branches based on the \lstinline{if} statement.
They are merged into one branch after executing assignment statements.
Every statement has semantics like this.
The whole process can be expressed in an automaton that is a concatenation of the automata of its statements.

\subsubsection{\lstinline{send} and \lstinline{recv}}

\begin{figure}
  \centering
  \includegraphics[scale=0.6]{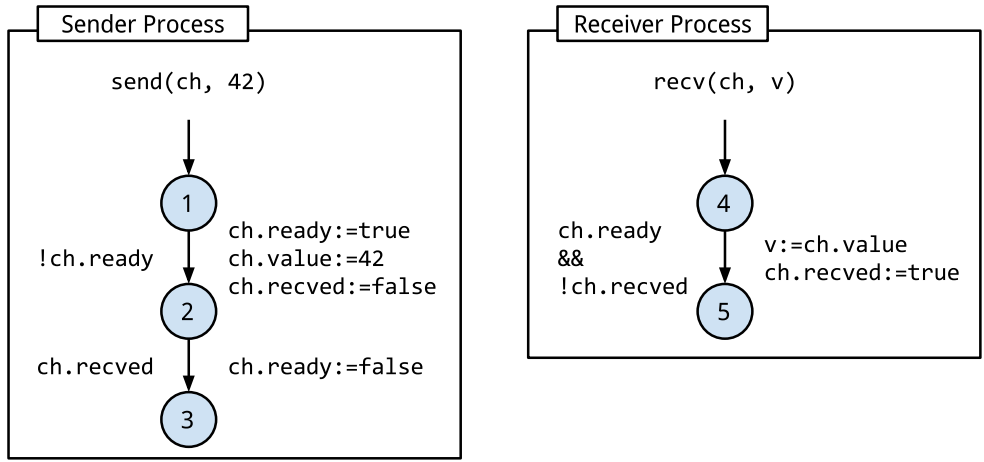}
  \caption{Two Processes Exchanging a Value}
  \label{send-recv-statements.pdf}
\end{figure}

A pair of \lstinline{send} and \lstinline{recv} statements cooperate to exchange messages. 
Figure~\ref{send-recv-statements.pdf} shows the semantics of them. 
To show the process of this exchange, consider two processes trying to send and receive a value via a rendezvous channel. Sending a value via a rendezvous channel has been done by using three internal variables in a channel: a ready flag, a received flag and a value buffer. The initial states of flags are false. The procedure follows.

\begin{enumerate}
\item In the initial setting, two processes are at the state 1 and the state 4 (in Figure~\ref{send-recv-statements.pdf}). The initial value of the ready flag is false. 
Thus the sender process can proceed to the state 2 while the receiver process cannot proceed to the state 5.
\item After the sender process steps to the state 2, the ready flag is true and the value to be sent is set to the buffer. At this point, the sender process is blocked because the received flag is false and the receiver process can make a step to the state 5. The receiver process receives a value from the buffer and set the received flag.
\item The sender process can proceed to state3, and the whole exchange process has been completed.
\end{enumerate}

With this process, only one sender and receiver can communicate in a channel at once, and, even if one process tries to communicate, it blocks until the other process comes to communicate with it.

\subsubsection{\lstinline{timeout_recv} and \lstinline{nonblock_recv}}

Sandal provides two variants of \lstinline{recv} statement: \lstinline{timeout_recv} and \lstinline{nonblock_recv}.
Unlike \lstinline{recv} statement, 
they are provided as functions because they should return boolean values that express the status of timeout.

\begin{figure}
  \centering
  \includegraphics[scale=0.6]{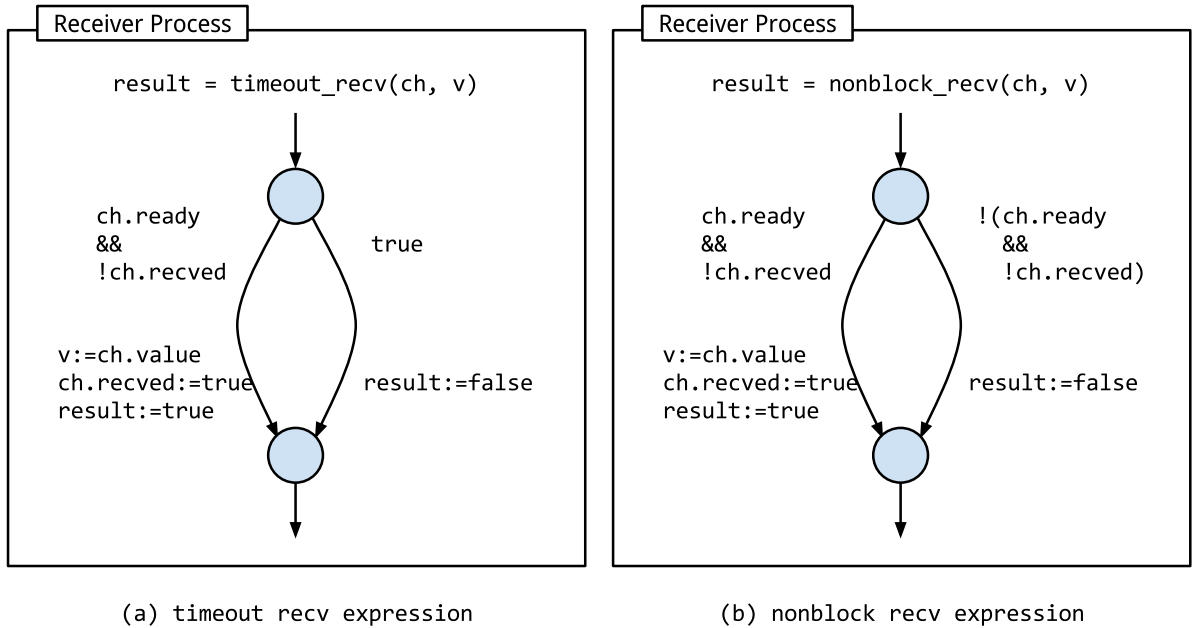}
  \caption{\lstinline|timeout_recv| and \lstinline|nonblock_recv|}
  \label{timeout-recv-and-nonblock-recv.pdf}
\end{figure}

A \lstinline{timeout_recv} expression receives a value from a specified channel as \lstinline{recv} statement.
In addition, it may perform a timeout action modeled as the right branch shown in Figure~\ref{timeout-recv-and-nonblock-recv.pdf} (a). 
If a value is successfully received (in Figure~\ref{timeout-recv-and-nonblock-recv.pdf}), the expression evaluates to true.
Otherwise, it evaluates to false.

The expression may perform timeout action even if the corresponding sender process is ready to send a value. 
This is an intended behavior. Since network delay is unbound, the communication always has a chance to unable to complete a transmission in a specific time window. The behavior of \lstinline{timeout_recv} expressions reflects these cases.

A \lstinline{nonblock_recv} expression is another variant of \lstinline{recv} statement.  
It receives a value only if it is ready. 
The behavior is modeled with two branches as shown in Figure~\ref{timeout-recv-and-nonblock-recv.pdf} (b).
Only one branch can be chosen since the guard of one branch is the negation of the other.
If a value is ready and is successfully received, the expression itself evaluates to true.
Otherwise, the expression evaluates to false.

The difference between \lstinline{timeout_recv} and \lstinline{nonblock_recv} is that 
the former is categorized as a fault.
As Sandal injects faults in a non-deterministic manner, the timeout fault will be injected non-deterministically. 
On the other hand, \lstinline{nonblock_recv} is not a fault and the additional behavior is not added in a non-deterministic manner.

\subsubsection{Unexpected Termination of Processes}

Unexpected termination of processes is a fault that a process is unintentionally shutdown. 
Using this fault, we can express machine crashes or process crashes.
This type of faults will be injected to processes that have \lstinline|@shutdown| fault markers.

\begin{figure}
    \centering
    \includegraphics[scale=.6]{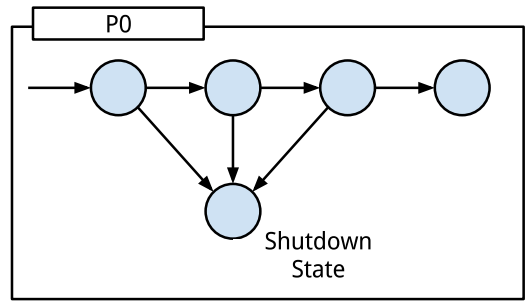}
    \caption{Unexpected Terminations of Processes}
    \label{process-termination.pdf}
\end{figure}

To implement this fault, a shutdown state is introduced and transitions to the state are added in the target process (Figure~\ref{process-termination.pdf}). 
These newly added transitions have no guard conditions nor actions. The transitions are injected before and after the execution of statements. This means that each statement is an atomic action, and the termination fault does not interfere with their execution.

This type of fault are also implemented in non-deterministic way; the chance to execute a statement normally and the chance to go to the shutdown state are even. Model checker tries both choices and tries all combination of these choices. By harnessing non-determinism, model checker can simulate arbitrary shutdown scenarios.

\begin{figure}
    \centering
    \includegraphics[scale=.6]{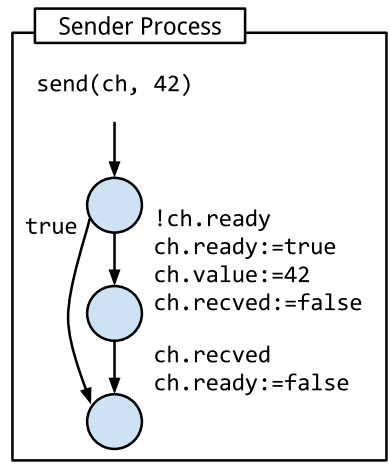}
    \caption{Random Loss of Messages}
    \label{random-loss.pdf}
\end{figure}

\subsubsection{Random Loss of Messages}

Random loss of messages is a fault that some messages are dropped. 
Thus no receiver will be able to receive them. This type of faults will be injected to channels that have \lstinline|@drop| fault markers, and all \lstinline{send} statements over those channels start to drop a message occasionally.

The implementation of this fault is done by modifying the semantics of \lstinline{send} statements of those faulty channels. The modified \lstinline{send} statements may skip their normal behavior occasionally (Figure~\ref{random-loss.pdf}).

\section{Case study}
\label{section:CaseStudy}

As a case study of this work, this section show the modeling and verification of two-phase commit protocol that is an algorithm to solve a consensus problem. It provides a way to determine a value which is acknowledged by all of the machine participated. It is used in major database systems such as MySQL to realize a transaction over multiple nodes.

\begin{figure}
  \centering
  \includegraphics[width=.8\linewidth]{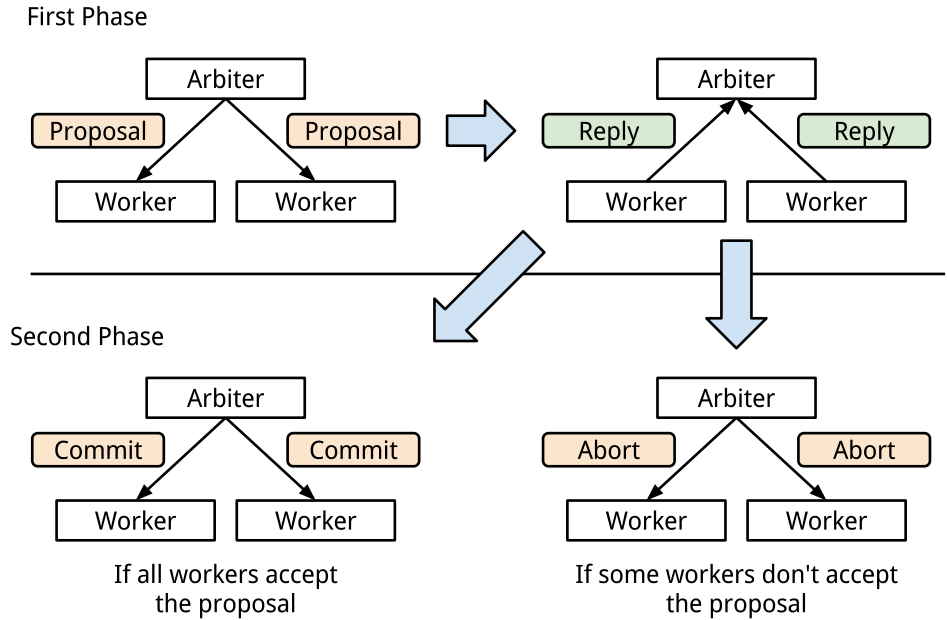}
  \caption{Two-Phase Commit Protocol}
  \label{two-phase-commit.pdf}
\end{figure}

The algorithm is performed by a single process called an arbiter and two or more processes called workers (Figure~\ref{two-phase-commit.pdf}). The arbiter initiates the protocol and proposes a value. The workers receive requests from the arbiter and send replies to it. In the first phase of the protocol, the arbiter sends a proposal to the workers. Each worker checks the proposed value and replies whether it is acceptable or not. In the second phase, the arbiter aggregates the replies from the workers and see if all of the workers can accept the proposed value. If the value is acceptable, the arbiter sends a commit message to the workers. The workers received a commit message should accept the proposed value. If one worker replies the value is not acceptable in the first phase, the proposal fails, and the arbiter sends abort messages to the other workers.

\begin{figure}
\begin{lstlisting}[basicstyle={\small\ttfamily},frame=lines]
data Response { Ready, NotReady, Commit, Abort }
proc Arbiter(chRecvs []channel { Response }, 
             chSends []channel { Response }) {
  var determined bool = false
  for ch in chSends {
    send(ch, Ready)
  }
  var all_ready bool = true
  for ch in chRecvs {
    var resp Response
    var recved bool = timeout_recv(ch, resp)
    if !recved || (recved && resp != Ready) {
      all_ready = false
    }
  }
  determined = true
  if all_ready {
    for ch in chSends {
      send(ch, Commit)
    }
  } else {
    for ch in chSends {
      send(ch, Abort)
    }
  }
}
proc Worker(chRecv channel { Response }, chSend channel { Response }) {
  var resp Response
  recv(chRecv, resp)
  choice { send(chSend, NotReady) }, { send(chSend, Ready) }
  recv(chRecv, resp)
}
init {
  chWorker1Send : channel { Response } @drop,
  chWorker1Recv : channel { Response } @drop,
  chWorker2Send : channel { Response } @drop,
  chWorker2Recv : channel { Response } @drop,
  arbiter : Arbiter([chWorker1Send, chWorker2Send],
                    [chWorker1Recv, chWorker2Recv]) @shutdown,
  worker1 : Worker(chWorker1Recv, chWorker1Send) @shutdown,
  worker2 : Worker(chWorker2Recv, chWorker2Send) @shutdown,
}
ltl {
  F (G (arbiter.determined &&
     ((!arbiter.all_ready) -> 
        (!(worker1.resp == Commit) && !(worker2.resp == Commit)))))
}
\end{lstlisting}
\caption{A Two-Phase Commit Model in Sandal}\label{two-phase-commit.sandal}
\end{figure}

In this case study, several models are written in Sandal and Promela. 
One is the model of two-phase commit protocol without any fault, and the rest is ones with faults. 
The injected faults are random loss of messages, unexpected termination of processes, and timeout in receiving messages. In each model, the safety property of two-phase commit is verified. A Sandal model with these faults is shown in Figure~\ref{two-phase-commit.sandal}. Random loss of messages and unexpected termination of processes are injected by adding fault markers and timeout in receiving messages is injected by replacing recv statements of the arbiter with timeout recv statements.

The verification results show both the Sandal models and the Promela models that produce the valid results; the safety property holds for the model without faults and the model with timeout faults. The reason that the property holds with timeout faults is because the models fallback to the abort behavior if the arbiter cannot receive workers' replies. The safety property does not hold for the model with random loss of messages and the model with unexpected termination of processes.

The sizes of the models are measured by lines. Table~\ref{two-phase-commit-size} shows lines of the eight models. Since Sandal has a built-in fault support, the sizes of the models do not grow even if some faults are injected, and the models do not lose their maintainability. The injected faults are well controlled by the language so that unwanted side-effects do not occur. The Promela models increase their sizes as some faults are injected. To overcome this issue, an automatic fault-injection tool is needed, but avoiding unwanted side-effects is still hard to accomplish.

Aside from the validity of the verification, the verification speed is also a matter of concern. The benchmark is taken using Linux 3.12.9 running on top of a PC with Intel Core i7-3770K 3.50GHz and 16GB memory.
For model checking, we use NuSMV 2.5.4 (as the backend of Sandal) and Spin 6.2.5. 
The execution times needed to verify the models are shown in Table~\ref{two-phase-commit-speed}. 
It shows the verification speed of Sandal is still acceptable even if some faults are injected. It is interesting that the speed is increased or decreased when injecting faults in Sandal while there are no differences among the speeds of the Spin models. The reason for this is considered to be the difference of the model checking algorithms. 
NuSMV, the backend of Sandal, does symbolic model checking while Spin does explicit model checking. 

The resources consumed in the verifications are shown in Table~\ref{two-phase-commit-bdd} and Table~\ref{two-phase-commit-state}.
They show the number of the BDD nodes allocated by NuSMV and the number of the states stored by Spin respectively.

No significant relationships between the BDD node sizes and the verification speeds can be observed. 
The number of states in some Promela models are very small.
This is because Spin stops verification when it find the first counter-example.

\begin{table}
  \begin{tabular}{cc}
    \begin{minipage}{.49\hsize}
      \begin{center}
        \caption{Sizes of 2PC Models}
        \begin{tabular}{l|c|c}
          \hline
                            & Sandal   & Promela  \\
          \hline \hline
          No fault          & 51 lines & 66 lines \\
          With Timeout      & 51 lines & 74 lines \\
          With Message loss & 51 lines & 70 lines \\
          With Termination  & 51 lines & 98 lines \\
          \hline
        \end{tabular}
        \label{two-phase-commit-size}
      \end{center}
    \end{minipage}

    \begin{minipage}{.49\hsize}
      \begin{center}
        \caption{Verification Speeds of 2PC Models}
        \begin{tabular}{l|c|c}
          \hline
                            & Sandal   & Spin     \\
          \hline \hline
          No fault          & 0.96 sec & 1.01 sec \\
          With Timeout      & 2.88 sec & 1.02 sec \\
          With Message loss & 2.11 sec & 1.06 sec \\
          With Termination  & 0.51 sec & 1.17 sec \\
          \hline
        \end{tabular}
        \label{two-phase-commit-speed}
      \end{center}
    \end{minipage}

    \\
    \\

    \begin{minipage}{.49\hsize}
      \begin{center}
        \caption{Allocated BDD Nodes}
        \begin{tabular}{l|c}
          \hline
          No fault          & 925483 \\
          With Timeout      & 261547 \\
          With Message loss & 369272 \\
          With Termination  & 588751 \\
          \hline
        \end{tabular}
        \label{two-phase-commit-bdd}
      \end{center}
    \end{minipage}

    \begin{minipage}{.49\hsize}
      \begin{center}
        \caption{States Stored}
        \begin{tabular}{l|c}
          \hline
          No fault          & 110 states \\
          With Timeout      & 305 states \\
          With Message loss & 11 states  \\
          With Termination  & 7 states \\
          \hline
        \end{tabular}
        \label{two-phase-commit-state}
      \end{center}
    \end{minipage}

  \end{tabular}
\end{table}

\section{Future Work}\label{section:FutureWork}

The result reported in this paper is a part of our ongoing work towards the verifiable framework for self-adaptable distributed systems based on a reflective architecture proposed by the second author\cite{Watanabe:2013aa}.
To achieve the goal, we need to establish a modeling framework for general adaptable (or reflective) behaviors
The current version of Sandal only provides limited features for modeling adaptable behaviors (a fixed set of fault-handling actions) as its language constructs.
Based on this work, future work is discussed as follows.

\paragraph{Fault as a Cross-Cutting Concern}
A cross-cutting concern is a feature that affects multiple parts of the program.  These concerns include authentication and logging. Because they are difficult to compose in a modular way in many cases, their implementation is scattered in the source code of the program. Therefore, separation of concerns principle is often violated.

Aspect-oriented programming (AOP) \cite{Kiczales:1997aa} is one of the approaches for this problem. AOP is motivated to increase modularity by enforcing separation of concern principle. 

In the fault-injection approach proposed in this paper, faults can be captured as a cross-cutting concern. It affects multiple parts of a system, and the realization of a fault is scattered in a model. If the modeling language employ AOP, faults as well as other adaptable behaviors can be organized into aspects and incorporated into the model by a similar mechanism to aspect weaving. Actually aspect-oriented approach is proven to be effective not only in programming languages but also in modeling languages\cite{Yamada:2006aa}.

\paragraph{Feedback of Failure Detectors}
Failure detector is a mechanism that enables a machine to estimate failures in a system \cite{Chandra:1996aa}. The failure it can treat varies. Most simple one is estimating other machine's crashes. The timeout feature of Sandal is also one of failure detection. It can detect a message cannot be sent in some time window and can feedback to the system. Failure detectors do only estimation due to the limitation of the reliability of themselves. Besides this limitation, they contribute constructing fault-tolerant distributed systems.

Giving feedback from failure-detection mechanisms to system models is one of the future work. The mechanism itself is sometimes unrelated to a system. It can be abstracted away from the system model specification, and, thus, modeling languages can provide a way to describe those mechanisms with modularity.

\section{Concluding Remark}
\label{section:Conclusion}

We propose a linguistic approach to reducing the cost of modeling fault-prone distributed systems.
The key technology is a variation of software fault injection (SFI) applied to process models used for model checking.
We designed and implemented a modeling language Sandal that supports the specification of typical faults in message-passing systems.
Using the Sandal compiler, all possible faults specified in a model is automatically injected into the result that can be model checked by NuSMV.
The advantage of the method is demonstrated by specifying and verifying models of the two-phase commit protocol.

\section*{Acknowledgments}
This work is partly supported by JSPS KAKENHI Grant No. 24500033.

\bibliographystyle{eptcs}
\bibliography{modstar2014}

\begin{thebibliography}{10}
\providecommand{\bibitemdeclare}[2]{}
\providecommand{\surnamestart}{}
\providecommand{\surnameend}{}
\providecommand{\urlprefix}{Available at }
\providecommand{\url}[1]{\texttt{#1}}
\providecommand{\href}[2]{\texttt{#2}}
\providecommand{\urlalt}[2]{\href{#1}{#2}}
\providecommand{\doi}[1]{doi:\urlalt{http://dx.doi.org/#1}{#1}}
\providecommand{\bibinfo}[2]{#2}

\bibitemdeclare{article}{Bolognesi:1987aa}
\bibitem{Bolognesi:1987aa}
\bibinfo{author}{Tommaso \surnamestart Bolognesi\surnameend} \&
  \bibinfo{author}{Ed~\surnamestart Brinksma\surnameend}
  (\bibinfo{year}{1987}): \emph{\bibinfo{title}{Introduction to the {ISO}
  specification language {LOTOS}}}.
\newblock {\sl \bibinfo{journal}{Computer Networks and ISDN Systems}}
  \bibinfo{volume}{14}(\bibinfo{number}{1}), pp. \bibinfo{pages}{25--59},
  \doi{10.1016/0169-7552(87)90085-7}.

\bibitemdeclare{inproceedings}{Bozzano:2003aa}
\bibitem{Bozzano:2003aa}
\bibinfo{author}{Marco \surnamestart Bozzano\surnameend} \&
  \bibinfo{author}{Adolfo \surnamestart Villafiorita\surnameend}
  (\bibinfo{year}{2003}): \emph{\bibinfo{title}{Improving System Reliability
  via Model Checking: The {FSAP}/{NuSMV}-{SA} Safety Analysis Platform}}.
\newblock In: {\sl \bibinfo{booktitle}{Computer Safety, Reliability, and
  Security (SAFECOMP 2003)}}, {\sl \bibinfo{series}{Lecture Notes in Computer
  Science}} \bibinfo{volume}{2788}, \bibinfo{publisher}{Springer-Verlag}, pp.
  \bibinfo{pages}{49--62}, \doi{10.1007/978-3-540-39878-3_5}.

\bibitemdeclare{article}{Chandra:1996aa}
\bibitem{Chandra:1996aa}
\bibinfo{author}{Tushar~Deepak \surnamestart Chandra\surnameend} \&
  \bibinfo{author}{Sam \surnamestart Toueg\surnameend} (\bibinfo{year}{1996}):
  \emph{\bibinfo{title}{Unreliable Failure Detectors for Reliable Distributed
  Systems}}.
\newblock {\sl \bibinfo{journal}{Journal of the ACM}}
  \bibinfo{volume}{43}(\bibinfo{number}{2}), pp. \bibinfo{pages}{225--267},
  \doi{10.1145/226643.226647}.

\bibitemdeclare{inproceedings}{Cimatti:2002aa}
\bibitem{Cimatti:2002aa}
\bibinfo{author}{Alessandro \surnamestart Cimatti\surnameend},
  \bibinfo{author}{Edmund \surnamestart Clarke\surnameend},
  \bibinfo{author}{Enrico \surnamestart Giunchiglia\surnameend},
  \bibinfo{author}{Fausto \surnamestart Giunchiglia\surnameend},
  \bibinfo{author}{Marco \surnamestart Pistore\surnameend},
  \bibinfo{author}{Marco \surnamestart Roveri\surnameend},
  \bibinfo{author}{Roberto \surnamestart Sebastiani\surnameend} \&
  \bibinfo{author}{Armando \surnamestart Tacchella\surnameend}
  (\bibinfo{year}{2002}): \emph{\bibinfo{title}{{NuSMV} 2: An OpenSource Tool
  for Symbolic Model Checking}}.
\newblock In: {\sl \bibinfo{booktitle}{Computer Aided Verification}}, {\sl
  \bibinfo{series}{Lecture Notes in Computer Science}} \bibinfo{volume}{2404},
  \bibinfo{publisher}{Springer-Verlag}, pp. \bibinfo{pages}{359--364},
  \doi{10.1007/3-540-45657-0_29}.

\bibitemdeclare{}{golang-web}
\bibitem{golang-web}
\emph{\bibinfo{title}{The {Go} Programming Language}}.
\newblock \urlprefix\url{http://golang.org}.

\bibitemdeclare{article}{Holzmann:1997aa}
\bibitem{Holzmann:1997aa}
\bibinfo{author}{Gerard~J. \surnamestart Holzmann\surnameend}
  (\bibinfo{year}{1997}): \emph{\bibinfo{title}{The Model Checker {Spin}}}.
\newblock {\sl \bibinfo{journal}{IEEE Transactions on Software Engineering}}
  \bibinfo{volume}{23}(\bibinfo{number}{5}), pp. \bibinfo{pages}{279--295},
  \doi{10.1109/32.588521}.

\bibitemdeclare{inproceedings}{Kiczales:1997aa}
\bibitem{Kiczales:1997aa}
\bibinfo{author}{Gregor \surnamestart Kiczales\surnameend},
  \bibinfo{author}{John \surnamestart Lamping\surnameend},
  \bibinfo{author}{Anurag \surnamestart Mendhekar\surnameend},
  \bibinfo{author}{Chris \surnamestart Maeda\surnameend},
  \bibinfo{author}{Cristina \surnamestart Lopes\surnameend},
  \bibinfo{author}{Jean-Marc \surnamestart Loingtier\surnameend} \&
  \bibinfo{author}{John \surnamestart Irwin\surnameend} (\bibinfo{year}{1997}):
  \emph{\bibinfo{title}{Aspect-Oriented Programming}}.
\newblock In: {\sl \bibinfo{booktitle}{ECOOP '97 -- Object-Oriented
  Programming}}, {\sl \bibinfo{series}{Lecture Notes in Computer Science}}
  \bibinfo{volume}{1241}, \bibinfo{publisher}{Springer-Verlag}, pp.
  \bibinfo{pages}{220--242}, \doi{10.1007/BFb0053381}.

\bibitemdeclare{article}{Sirjani:2004aa}
\bibitem{Sirjani:2004aa}
\bibinfo{author}{Marjan \surnamestart Sirjani\surnameend}, \bibinfo{author}{Ali
  \surnamestart Movaghar\surnameend}, \bibinfo{author}{Amin \surnamestart
  Shali\surnameend} \& \bibinfo{author}{Frank~S. \surnamestart {de
  Boer}\surnameend} (\bibinfo{year}{2004}): \emph{\bibinfo{title}{Modeling and
  Verification of Reactive Systems using {Rebeca}}}.
\newblock {\sl \bibinfo{journal}{Fundamenta Informaticae}}
  \bibinfo{volume}{63}(\bibinfo{number}{4}), pp. \bibinfo{pages}{385--410}.
\newblock
  \urlprefix\url{http://iospress.metapress.com/content/wg947keu129prhbd/}.

\bibitemdeclare{inproceedings}{Steiner:2004aa}
\bibitem{Steiner:2004aa}
\bibinfo{author}{Wilfried \surnamestart Steiner\surnameend},
  \bibinfo{author}{John \surnamestart Rushby\surnameend},
  \bibinfo{author}{Maria \surnamestart Sorea\surnameend} \&
  \bibinfo{author}{Holger \surnamestart Pfeifer\surnameend}
  (\bibinfo{year}{2004}): \emph{\bibinfo{title}{Model Checking a Fault-Tolerant
  Startup Algorithm: From Design Exploration To Exhaustive Fault Simulation}}.
\newblock In: {\sl \bibinfo{booktitle}{International Conference on Dependable
  Systems and Networks (DSN '04)}}, pp. \bibinfo{pages}{189--198},
  \doi{10.1109/DSN.2004.1311889}.

\bibitemdeclare{inproceedings}{Svenningsson:2010aa}
\bibitem{Svenningsson:2010aa}
\bibinfo{author}{Rickard \surnamestart Svenningsson\surnameend},
  \bibinfo{author}{Jonny \surnamestart Vinter\surnameend},
  \bibinfo{author}{Henrik \surnamestart Eriksson\surnameend} \&
  \bibinfo{author}{Martin \surnamestart T{\"o}rngren\surnameend}
  (\bibinfo{year}{2010}): \emph{\bibinfo{title}{{MODIFI}: A
  {MOD}el-{I}mplemented {F}ault {I}njection Tool}}.
\newblock In: {\sl \bibinfo{booktitle}{Computer Safety, Reliability, and
  Security}}, {\sl \bibinfo{series}{Lecture Notes in Computer Science}}
  \bibinfo{volume}{6351}, \bibinfo{publisher}{Springer-Verlag}, pp.
  \bibinfo{pages}{210--222}, \doi{10.1007/978-3-642-15651-9_16}.

\bibitemdeclare{inproceedings}{Watanabe:2013aa}
\bibitem{Watanabe:2013aa}
\bibinfo{author}{Takuo \surnamestart Watanabe\surnameend}
  (\bibinfo{year}{2013}): \emph{\bibinfo{title}{Towards a Compositional
  Reflective Architecture for Actor-Based Systems}}.
\newblock In: {\sl \bibinfo{booktitle}{Workshop on Programming based on Actors,
  Agents, and Decentralized Control (AGERE!@SPLASH 2013)}},
  \bibinfo{publisher}{ACM}, pp. \bibinfo{pages}{19--24},
  \doi{10.1145/2541329.2541341}.

\bibitemdeclare{inproceedings}{Yamada:2006aa}
\bibitem{Yamada:2006aa}
\bibinfo{author}{Kiyoshi \surnamestart Yamada\surnameend} \&
  \bibinfo{author}{Takuo \surnamestart Watanabe\surnameend}
  (\bibinfo{year}{2006}): \emph{\bibinfo{title}{An Aspect-Oriented Approach to
  Modular Behavioral Specification}}.
\newblock In: {\sl \bibinfo{booktitle}{Proceedings of 1st Workshop on
  Aspect-Based and Model-Based Separation of Concerns in Software Systems (ABMB
  2005)}}, {\sl \bibinfo{series}{Electronic Notes in Theoretical Computer
  Science}} \bibinfo{volume}{163(1)}, \bibinfo{publisher}{Elsevier}, pp.
  \bibinfo{pages}{45--56}, \doi{10.1016/j.entcs.2006.07.002}.

\end{thebibliography}

\end{document}